\documentclass[journal]{IEEEtran}

\usepackage[T1]{fontenc} 
\usepackage{amsmath}
\usepackage[cmintegrals]{newtxmath}
\usepackage{amsfonts} 
\usepackage{graphicx} 
\usepackage{braket}
\usepackage{hyperref}
\usepackage{color}
\usepackage{comment}
\usepackage{bm}
\usepackage{microtype}
\usepackage[noadjust]{cite}
\usepackage{gensymb}

\begin{document}
\bstctlcite{BSTcontrol}

\title{Modelling the Memory of Turbulence-Induced Beam Wander}
%
%
%

\author{Mitchell~A.~Cox,~\IEEEmembership{Member,~OSA,}
        Lucas~Gailele,
        Ling~Cheng,~\IEEEmembership{Senior Member,~IEEE,}
        and~Andrew~Forbes,~\IEEEmembership{Fellow,~OSA}
\thanks{M. Cox and L. Cheng are with the School of Electrical and Information Engineering, L. Gailele and A. Forbes are with the School of Physics, both at the University of the Witwatersrand, Johannesburg, South Africa.  mitchell.cox@wits.ac.za.}
\thanks{Manuscript received ...; revised ...}}



\maketitle

\begin{abstract}
One of the major challenges for long range, high speed Free-Space Optical (FSO) communication is turbulence induced beam wander. Beam wander causes fluctuations in the received intensity as well as crosstalk in mode division multiplexed systems. Existing models for beam wander make use of probability distributions and long term averages and are not able to accurately model time-dependent intensity fluctuations such as deep fading, where the received intensity is too low to maintain reliable communication for an extended period of time. In this work we present an elegant new memory model which models the behaviour of beam wander induced intensity fluctuations with the unique capability to accurately simulate deep fading. This is invaluable for the development of optimised error correction coding and digital signal processing in order to improve the throughput and reliability of FSO systems.
\end{abstract}

\begin{IEEEkeywords}
Free Space Optical Communication, Mode Division Multiplexing, Turbulence, Beam Wander
\end{IEEEkeywords}

%
\IEEEpeerreviewmaketitle

\section{Introduction}
\label{sec:intro}

Free-Space Optical (FSO) communication is an important alternative to both radio and optical fibre. Radio-based systems are often used to bridge short distances at relatively low cost such as ``the last mile'' and between neighbouring buildings. The problem of security, interference and consequently bandwidth has driven research in FSO communication due to its ability to address all three of these issues. More recently, Mode Division Multiplexing (MDM) has been used to demonstrate extremely high speed links over relatively short distances, but if the range could be improved then MDM technology may even become a viable alternative to high-speed fibre optic communication \cite{Willner2017, Zhao2015, lavery2018,Zhu2016,Qu2016,Krenn2016,Ren2016,Huang2014,Milione2015}. 

Over long distances a laser beam diverges, is attenuated by gasses and particles in the atmosphere and is distorted by atmospheric turbulence. Atmospheric turbulence can be described as two independent random processes, namely small and large scale \cite{Andrews2005}. The small scale effects of turbulence cause scintillation of a beam. This appears spatially as random deformations of intensity as well as phase aberrations of the beam wavefront which result in mode crosstalk - a significant problem for MDM. 

Large scale effects cause the beam itself to be deflected randomly as it propagates. This ``cork-screwing'' through the air eventually manifests as a lateral displacement at the receiver plane, where the beam randomly moves around the vicinity of the detection aperture (assuming reasonable alignment). Lateral displacement, in addition to angle of arrival fluctuations, manifest as beam wander at the focal (Fourier) plane of a receive aperture lens. Therefore, a small detector placed at this focal plane will measure fluctuations in intensity predominantly caused by beam wander, but also to a lesser extent by scintillation in the weak-to-moderate turbulence regime. 

In a communication system, this is called turbulence induced signal fading. There have been numerous studies into mitigating this using spatial modes, diversity and various signal processing techniques \cite{Willner2017, Chen2016a, Huang2018, Pang2018HGLG, cox2018, ndagano2017, cox2019hglg}, but turbulence remains a significant challenge. A key reason why turbulence is difficult to mitigate using signal processing and error correction coding is because the significant probability that the received intensity will drop below a usable threshold \textit{for an extended period of time}. The situation when the received intensity is too low for reliable communication is called a ``deep fade''. In addition to this, the channel has memory which causes the received intensity to remain in a certain state for a finite period of time. The combined effect of deep fading and memory is generally catastrophic to communication, where millions or even billions of bits can be lost.

This behaviour is not accounted for in existing models for turbulence induced fading. Existing fading models, for example the recent double generalised gamma-gamma model \cite{Kashani2015}, are able to predict the average time a channel will be in deep fade over the long term, but inherently cannot provide information about the channel memory, which in this case is the average duration of deep fade intervals. While techniques exist to deal with these so-called slow fading memory channels in power-line communications, for instance, they cannot be applied or tailored to FSO communications in the absence of a suitable channel model \cite{cheng2018fritchman}. 

In this work we propose a novel approach to modelling beam wander and beam wander induced fading, which is the dominant cause of fading in weak-to-moderate turbulence \cite{Andrews2005}. This approach is based on the hypothesis that the movement of the wandering beam is not a true random walk, but rather a random walk that has a level of correlation between successive position samples. To demonstrate this new approach, we experimentally measure beam wander over a 150~m link and present an Auto-Regressive Moving-Average (ARMA) based random walk modal which not only reproduces the results of existing models, but more importantly, allows for accurately simulating and determining the parameters of the channel memory. We show how this approach allows for the simulation of beam wander induced modal crosstalk using Orbital Angular Momentum (OAM) modes as an example. The unique ability of this approach to realistically simulate the evolution of modal crosstalk may facilitate the future development of robust FSO-MDM systems.

\section{Existing Beam Wander Models}
\label{sec:background}

In existing models, beam wander is spatially modelled by a Gaussian distribution which is characterised by a long-term average radial variance, $\braket{r_c^2}$, as well as the size of the received beam without any averaging (the short-term beam), denoted by $\omega_{\mathrm{ST}}$, illustrated in Fig.~\ref{fig:schematic} \cite{Andrews2005,Churnside1990,Rodriguez2005,esposito1967,fried1973wander,titterton1973}. It is intuitive to think that a smaller short-term beam which wanders around a receive aperture will result in a larger Gaussian shaped long-term beam, according to the central limit theorem \cite{Fante1975}. The size of the long-term beam, $\omega_\mathrm{LT}$, is given by 
\begin{equation}
    \omega_{\mathrm{LT}}^2 = \omega_{\mathrm{ST}}^2 + \braket{r_c^2}.
\end{equation}

Several analytical solutions for the beam wander radial variance exist. For infinite outer scale Kolmogorov turbulence, the beam wander radial variance is given by
\begin{equation}
    \label{eq:beamWanderVarInfOuterScale}
    \braket{r_c^2} = 2.42 C_n^2 L^3 \omega_0^{-1/3} {}_2F_1\!\left(1/3, 1; 4; 1 - |\Theta_0| \right),
\end{equation}
where $C_n^2$ is the refractive index structure constant which describes the strength of the turbulence, $L$ is the propagation distance, $\omega_0$ is the initial beam waist, ${}_2F_1$ is the hyper-geometric function and $\Theta_0$ is the beam parameter \cite{Andrews2005}. For a collimated beam ($\Theta_0 = 1$) the expression reduces to
\begin{equation}
    \label{eq:beamWanderVarInfOuterScaleColl}
    \braket{r_c^2} = 2.42 C_n^2 L^3 \omega_0^{-1/3},
\end{equation}
and for a finite outer scale turbulence and a collimated laser beam,  Eq.~\ref{eq:beamWanderVarInfOuterScale} becomes
\begin{equation}
    \label{eq:beamWanderVarFinOuterScaleColl}
    \braket{r_c^2} = 2.42 C_n^2 L^3 \omega_0^{-1/3} \left[ 1 - \left(\frac{\kappa_0^2 \omega_0^2}{1+\kappa_0^2\omega_0^2} \right)^{1/6} \right].
\end{equation}

In addition to the information provided by long-term averages and variances, it is often a requirement to perform Monte-Carlo simulations which require short-term information. In a well aligned optical communication system, the receiver photodiode would ideally be situated at the centre of the long term beam. By approximating the received beam as Gaussian with size $\omega_{ST}$, the received intensity at a moment in time when the beam wanders to a random radius is given by
\begin{equation}
    \label{eq:mapping}
    I_t(r) = I_0 \exp\!\!\left[-2\frac{r_{c,t}^2}{\omega^2_{ST}} \right],
\end{equation}
where $I_0$ is the intensity at the centre of the beam and as before, the random variable $r_c = N(0,\braket{r_c^2})$. The position of the short-term beam can instead be represented in Cartesian rather than Polar coordinates, and the angle can be ignored due to symmetry:
\begin{equation}
    \label{eq:polarCart}
    r_{c,t}^2 = \beta_{x,t}^2 + \beta_{y,t}^2
\end{equation}
where $\beta_{x,t}$ and $\beta_{y,t}$ are Cartesian coordinates for beam wander. The use of Cartesian coordinates is required for reasons important to the new modelling approach proposed in this work. As an aside, the effects of scintillation may be incorporated into this beam wander induced fading model by treating $I_0$ as a random variable representing scintillation \cite{dios2004,Andrews2007Wander}.

The Probability Distribution Function (PDF) of the received intensity, $I$, due to beam wander, as measured at the centre of the long-term beam is given by 
\begin{equation}
\label{eq:beamWanderFadingPDF}
p(I) = \gamma I^{\gamma-1} \quad \mathrm{where} \quad 0 \leq I \leq 1,
\end{equation}
where $\gamma$ is a parameter related to the ratio of beam wander displacement and the receiver size \cite{Kiasaleh1994}. 
While these expressions are extremely useful, they are limited by a lack of temporal information about the evolution of the beam wander: there is no correlation from one moment to the next as $r_c$ is simply a normal random variable. Consequently, while it is indeed possible to determine the overall percentage of time a link may be in deep fade, for instance, it is impossible to determine the average duration of the deep fades. This information is required to inform the design of optimised Forward Error Correction (FEC) code rates and lengths, interleaving depth and other techniques used to make a link more robust to atmospheric turbulence and in its absence results in over- or under-engineered systems based on potentially incorrect assumptions.

\section{Methods}
\label{sec:methods}

\subsection{Experimental Setup}
\label{sec:setup}

\begin{figure}[tb]
    \centering
	\includegraphics[width=1\linewidth]{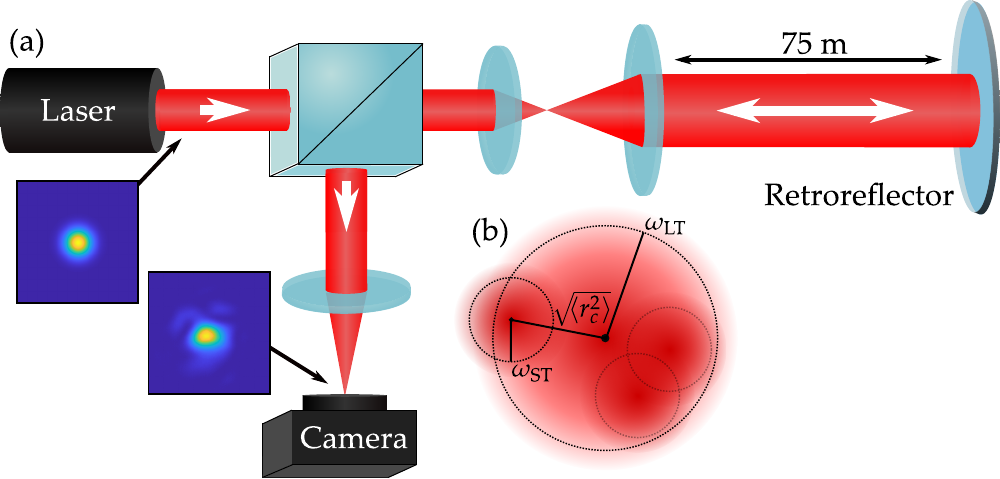}
	\caption{\label{fig:schematic} (a) Simplified schematic diagram of the experimental setup with insets of the beam intensity at the transmitter and receiver. (b) A diagram that illustrates the relationship between short- and long-term beam wander parameters.}
\end{figure}

Empirical measurements of beam wander were performed on a $L=150$~m horizontal outdoor link that was built at the CSIR in Pretoria, South Africa, which is at an altitude of 1.4~km. The link was approximately 5~m above the ground and used a 635~nm collimated laser beam with a diameter of approximately $\omega_0 = 7$~mm at the transmit aperture. A simplified diagram of the setup is shown in Fig.~\ref{fig:schematic}~(a). The transmitted and received laser beams were expanded and reduced by a telescope and imaged onto a high speed camera at the Fourier plane. The received beam was measured by a camera to resolve the spatial movement of the beam over approximately ten seconds. 

The frame rate of the camera is an important parameter in this experiment as the movement of the beam must be recorded without loss of information. The Greenwood frequency, $f_G$, is a measure of the rate at which turbulence affects a beam and is easily calculated given the wind speed and atmospheric coherence length, $r_0$, as $f_G = 0.43(V/r_0)$ in Hertz \cite{Greenwood1977}. The weather at the time of measurement was clear, 28~$\degree$C with an average wind speed of $V=10$~km/h with gusts of up to $V=12$~km/h. With preliminary measurements of $r_0$ and assuming a constant but worst-case wind speed over the the course of a measurement, the Greenwood frequency was approximately $f_G = 130$~Hz. The camera frame rate was therefore conservatively set at 300~Frames Per Second (FPS) to capture the movement of the beam without loss of temporal information.

The position of the beam in each frame is found using the weighted centroid and stored as separate time series of $x$ and $y$ coordinates. Since the atmosphere is isotropic and the axes are orthogonal and thus independent, we restrict analysis to a single axis for convenience. When both axes are required for simulation of the resulting beam wander model, we simply run two models independently.

\subsection{ARMA-based Memory Model}
\label{sec:arima}

There are several classes of model that are suitable for correlated time series, depending on the characteristics of the signal \cite{box2015time,Hochreiter1997,Baum1966}. For stationary signals such as beam wander a so-called ARIMA (Auto-Regressive Integrated Moving Average) model is elegantly suitable, as evidenced by the results presented in Sec.~\ref{sec:fitting}, and can be fit using the Box-Jenkins method \cite{box2015time}. Auto-Correlation and Partial Auto-Correlation Functions (ACFs and PACFs) are used to approximately determine the order and suitability of the various model parameters, but ultimately the model fits performed using least-squares regression are verified by minimising the Akaike Information Criterion (AIC) and Bayesian Information Criterion (BIC).

A long term trend in the data is not expected, since a constant turbulence strength is assumed over the short duration of the measurements. Consequently, an ARMA($p$,$q$) model can be used without any Integral terms, where in general the position of the current sample is given by
\begin{equation}
    \label{eq:arma}
    \beta_{x,t} = c + \sum_{i=1}^p M_i \beta_{x,t-i} + \sum_{j=1}^q N_j \epsilon_{t-j} + \epsilon_t,
\end{equation}
where $c$ is a constant mean term, $p$ is the order of the auto-regressive series, $q$ is the order of the moving average series and $M_i$ and $N_j$ are the model factors of their respective series. Randomness is introduced to the model by the addition of $\epsilon_t$ which is zero mean Gaussian white noise with a suitable variance, $\sigma^2$. 

As Eq.~\ref{eq:arma} is directly modelling the position of the beam at a specific time, $\beta_{x,t}$, we are conveniently able to map this to a receive intensity using Eq.~\ref{eq:mapping} and \ref{eq:polarCart}. The required $\beta_{y,t}$ values are found using an identical but independent model since the atmosphere is isotropic.

\section{Fitting the Memory Model}
\label{sec:fitting}

To demonstrate the feasibility, effectiveness and advantages of the proposed modelling approach, we present a model for the link described in Sec.~\ref{sec:setup} under moderate-to-strong turbulence conditions where deep fading is more likely than in weak turbulence.

\begin{figure}[t]
    \centering
	\includegraphics[width=1\linewidth]{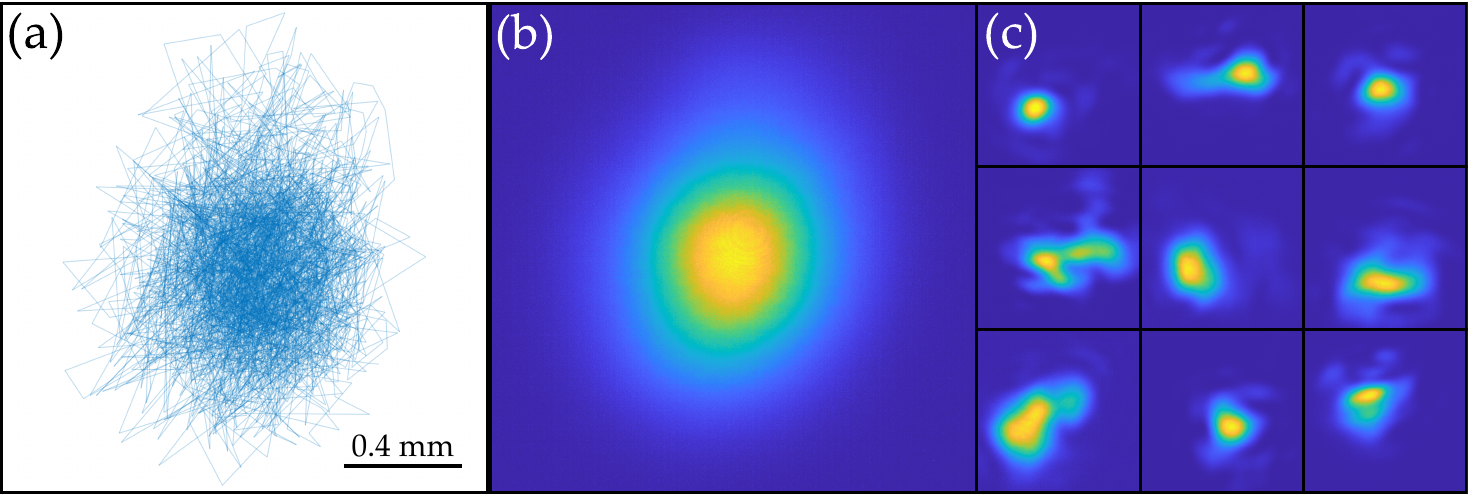}
	\caption{\label{fig:beams} Experimental measurements of the beam wander showing a plot of the centroid position in (a), the long-term beam in (b) and sample short term beam images in (c). }
\end{figure}

The turbulence parameters at the time of measurement are as follows, with images of the long- and short-term beams as well as the path that the short term beam took over the 10~s measurement window shown in Fig.~\ref{fig:beams}. The average refractive index structure constant $C_n^2 = 4.1\times10^{-13}$, the atmospheric coherence length $r_0 = 0.01$~m and the scintillation index is $\sigma_I = 0.55$, indicating that the link is in the weak to moderate turbulence regime. 

\begin{figure}[t]
    \centering
	\includegraphics[width=1\linewidth]{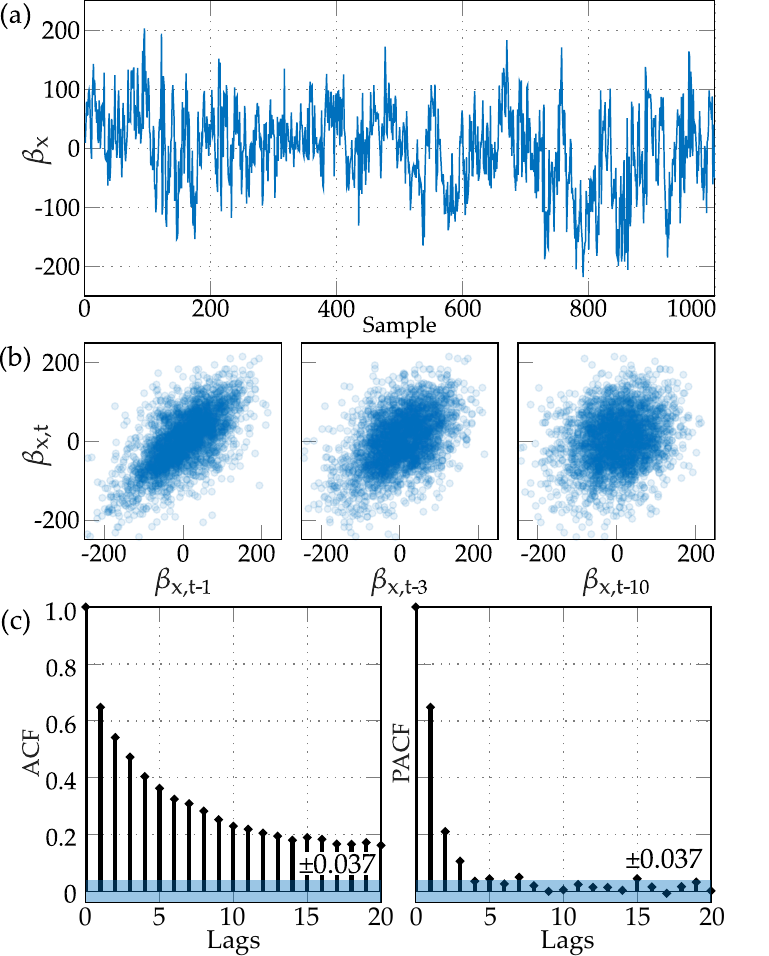}
	\caption{\label{fig:wanderDetail} (a) A short sample of the beam wander in $x$. (b) Scatter plots showing temporal correlation between samples at $t-1$, some correlation at $t-3$ but no correlation at $t-10$. (c) Auto-correlation and partial auto-correlation plots of the data with the 95\% significance region shaded ($\pm 0.037$). }
\end{figure}

One thousand samples of the change in centroid position over time, focusing on $\beta_{x,t}$, are shown in Fig.~\ref{fig:wanderDetail}~(a). Note that while we refer to discrete samples of $t$, the sample period is in fact 3.3~ms. The hypothesised temporal correlation between subsequent samples is made clear by simply plotting $\beta_{x,t}$ against the previous position, $\beta_{x,t-1}$, shown in Fig.~\ref{fig:wanderDetail}~(b). It is clear that there is a correlation with sample $t-1$ and also a correlation at sample $t-3$, however, at a lag of $t-10$ there is no obvious correlation.  The ACF and PACFs of the beam position are a more rigorous measure for this and are shown in Fig.~\ref{fig:wanderDetail}~(c). Clearly, a strong correlation exists for multiple samples and hence several milli-seconds. We can use this observation to build a model of the beam wander that is not simply a random walk as historically assumed, but rather a time series with temporal structure.

\begin{figure}[tb]
    \centering
	\includegraphics[width=1\linewidth]{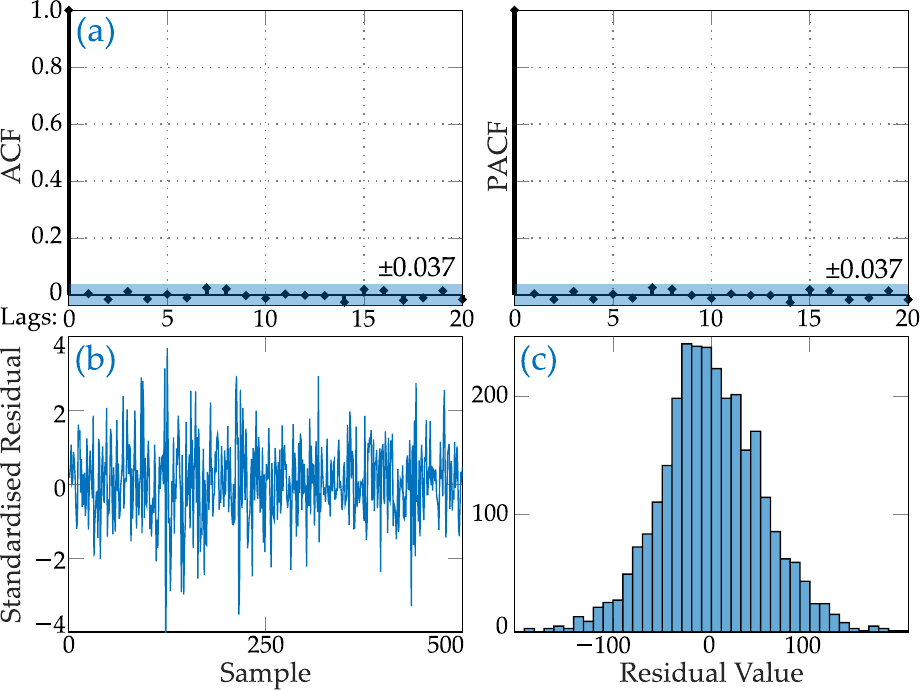}
	\caption{\label{fig:statsCheck} Plots to show that the residuals of the model and the measured samples are white Gaussian noise. (a) The ACF and PACF indicate no correlations. (b) A short sample of the standardised residuals showing no visible pattern. (c) The histogram of the residuals has a random Gaussian characteristic shape.  }
\end{figure}

It was found that $\beta_{x,t}$ (and by nature $\beta_{y,t}$) is well modelled by an ARMA(2,2) process with the model factors shown in Tab.~\ref{tab:arma}, which were found using a least squares regression on the approximately 3000 measurement samples. The order of the model was verified by testing orders from zero to twenty and simultaneously minimising the AIC and BIC criterion's, as per Sec.~\ref{sec:arima}. Another test of the resulting model is given by the statistics of the residuals, which should resemble random noise. Some of the residual statistics are visualised in Fig.~\ref{fig:statsCheck}. It is clear from the figure that there is no correlation between subsequent residual samples and a histogram of the residual values has a Gaussian shape, implying that the residuals are Gaussian white noise. The proposed ARMA(2,2) model is an excellent fit of the measured data in a theoretical sense, but is it a good fit in the physical sense and how can we use it to our advantage?

\begin{table}[tb]
\centering
\caption{Weights for the ARMA(2,2) model. As expected, the constant term is zero.}
\begin{tabular}{cccccc}
 $c$ & $M_1$ & $M_2$ & $N_1$ & $N_2$ & $\sigma^2$  \\
 \hline
 0 & 1.759 & -0.7626 & -1.289 & 0.3166 & 2150 \\
\end{tabular}
  \label{tab:arma}
\end{table}

\section{Modelling Beam-Wander Induced Fading}
\label{sec:discussion}

\begin{figure}[tb]
    \centering
	\includegraphics[width=1\linewidth]{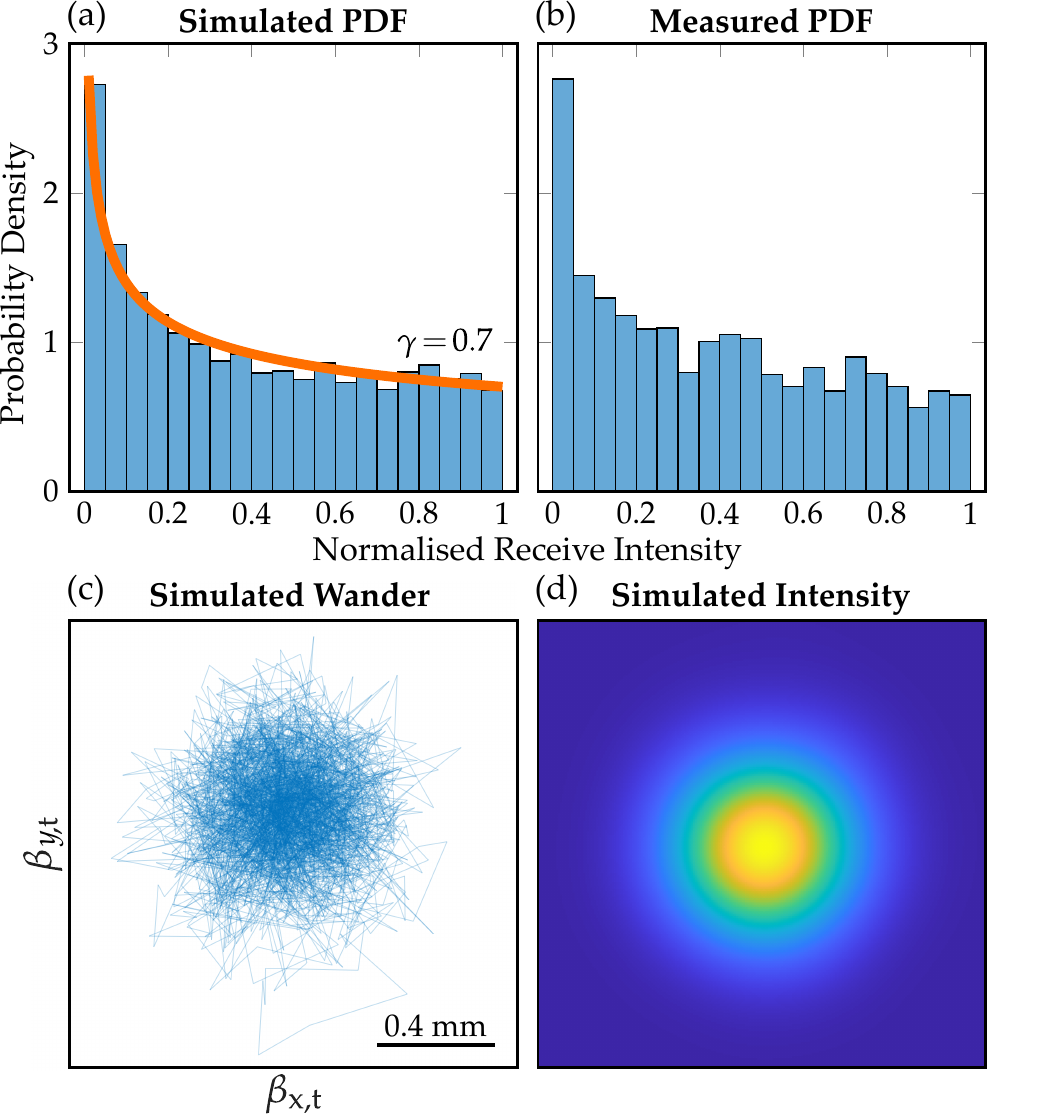}
	\caption{\label{fig:simWander} Plots to demonstrate the fading statistics of the model in comparison to experimental data. (a) and (b) are PDFs of simulated and measured intensities with an overlay (orange) of the theoretical PDF showing excellent agreement with existing theory. (c) Simulated beam wander positions with the resulting long-term intensity in (d).}
\end{figure}

In order to verify this new model, and hence this new approach, we must show that it is able to reproduce the results of existing models for beam wander. For a fair statistical comparison to the experimental data, the same number of samples of the ARMA(2,2) model presented in Sec.~\ref{sec:fitting} are generated as the number of experimental data samples. Plots of the results in this section are shown in Fig.~\ref{fig:simWander}. 

As discussed in Sec.~\ref{sec:background}, it is possible to calculate a simulated long-term intensity by summing the short-term beams over all of the simulated position samples. The resulting long term intensity distribution should have a Gaussian shape with a radial variance that matches that of the experimental data. The radial variance of the measured data gives $\braket{r_c^2}=4.31\times10^{-4}$ and for the simulated data $\braket{r_c^2}=4.26\times10^{-4}$. According to theory in Eq.~\ref{eq:beamWanderVarInfOuterScaleColl} for the estimated $C_n^2$ of the experiment, $\braket{r_c^2}=7.10\times10^{-4}$. The simulated and measured beam wander radial variances are very similar whereas the theory value is in the same order of magnitude but slightly different. This is due to the fact that the theory used does not take outer-scale effects into account because the outer scale was not measured. 

In addition to the long term intensity, by applying Eq.~\ref{eq:mapping} to the simulated and measured data, a comparison between the fading PDFs of the two data sets can be made. For a valid model, the shape of the simulated PDF must be similar to that of the measured data. Upon examination of Fig.~\ref{fig:simWander}, we see that this is indeed the case. It is clear that this new approach results in models that accurately reduce to existing models for beam wander, however, their real benefit is the ability to model the channels memory.

\section{Modelling Memory and Deep Fading}
\label{sec:memory}

As discussed in Sec.~\ref{sec:intro} and \ref{sec:background}, an important property in the design of a FSO link with memory is the average length of time the received intensity spends above or below a certain threshold. If the threshold is set to a low value that represents a deep fade for example, one can estimate the expected average duration of a deep fade and optimally design the system. This complements the conventional metric that is the expected probability that the link will be in deep fade, which is typically used to calculate channel capacity. 

Rather than referring to the length of time on either side of a threshold in seconds, for example, we instead refer to the so-called ``transition run length'' which is the number of consecutive samples on either side of the threshold. If required, this can be converted to time by multiplying by the sample period. This terminology is typically used for memory channels, for instance those described by Gilbert-Elliot Markov processes \cite{cheng2018fritchman}.

A Run Length Distribution (RLD) is a histogram of the number of occurrences of a certain transition run length above or below the specified threshold. Logically, with more data, longer transition run lengths (which have correspondingly lower probabilities) will begin to appear. For this reason, we can only compare data sets with the same number of samples. Due to the limited number of experimental samples available, we set the transition threshold to the mean intensity value for illustrative purposes. 

\begin{figure}[tb]
    \centering
	\includegraphics[width=1\linewidth]{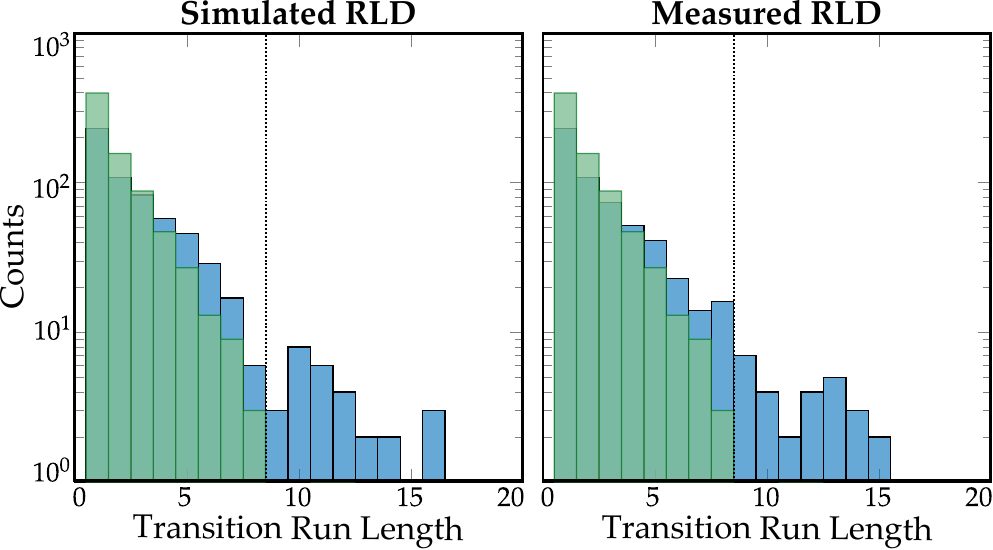}
	\caption{\label{fig:runLength} Run length distribution plots of the simulated ARMA model and the experimental measurements with a transition threshold set at 50\% intensity for illustrative purposes. Overlaid on the plots in green is the run length distribution of data simulated using the conventional PDF-based model which clearly does not match the experimental measurements.}
\end{figure}

Figure~\ref{fig:runLength} shows run length distribution plots for the ARMA simulated and experimentally measured beam wander induced fading data. The simulated data matches the experimental data well. Superimposed on the figure is a run length distribution plot for simulated data based on the conventional PDF-based model for beam wander (see Eq.~\ref{eq:beamWanderFadingPDF} with $\gamma=0.7$, which is shown in Fig.~\ref{fig:simWander}). The run length distribution of the conventional PDF-based model drops off steeply and is clearly not a match for the experimental data.

Run lengths of longer than eight consecutive samples above or below the threshold are rare when using the existing PDF model when using only 3000 samples. Obviously, if a larger number of random samples were generated, longer run lengths would be measured, but this would also be the case with more experimental data. In comparison with the measured data, longer run lengths of up to 15 consecutive samples are present even in a small data set. 

This is a clear deficiency of a PDF-based model: it is unlikely that any PDF which simultaneously matches both the PDF of the received intensity and of the run length distribution can be found. In contrast, the model produced by the approach proposed in this work accurately matches both the measured run length distribution as well as the fading distribution with only a marginal increase in complexity.

\section{Modelling Mode Crosstalk}

The proposed model has further utility than modelling received intensity fluctuations and their memory. Reliable and high speed long distance MDM systems are extremely challenging because of crosstalk issues. The computational complexity of strong error correction codes and MIMO is prohibitive at link speeds requiring the use of MDM in the first place. One of the dominant causes of crosstalk error in MDM systems is lateral displacement and tip / tilt at the receiver \cite{Lin2010oamtilt,Vasnetsov2005,cox2019hglg}. Alternatively, in modal diversity systems, modal crosstalk may be advantageous, in a sense, and can be used to design longer distance or more robust FSO systems \cite{Stevens2016,Geisler2018,cox2018}. Using the modelling approach in this paper it becomes possible to simulate and even predict the temporal evolution of mode crosstalk due to lateral displacement and hence facilitate the creation of optimised (and possibly less computationally intensive) schemes to correct errors due to crosstalk in both MDM and modal diversity. 

Here we briefly demonstrate this concept using Laguerre-Gaussian (LG) mode crosstalk. A transmitted LG mode with $\ell=0$, which is a standard Gaussian, will result in a spectrum of detected OAM modes at the receiver. Similarly, a transmitted higher-order mode in the presence of lateral displacement would result in a different spectrum of received modes.

This is often thought of simply as misalignment, where it is a challenge to efficiently couple the received light into a single mode optical fibre. In essence, it is for this reason that the use of few- or even multi-mode fibres (which can accept neighbouring OAM modes) has been demonstrated to be advantageous, especially when combined with digital signal processing \cite{Geisler2018}. A knowledge of the expected number of modes in this system could inform the choice of few-mode fibre. 

\begin{figure}[t]
    \centering
	\includegraphics[width=1\linewidth]{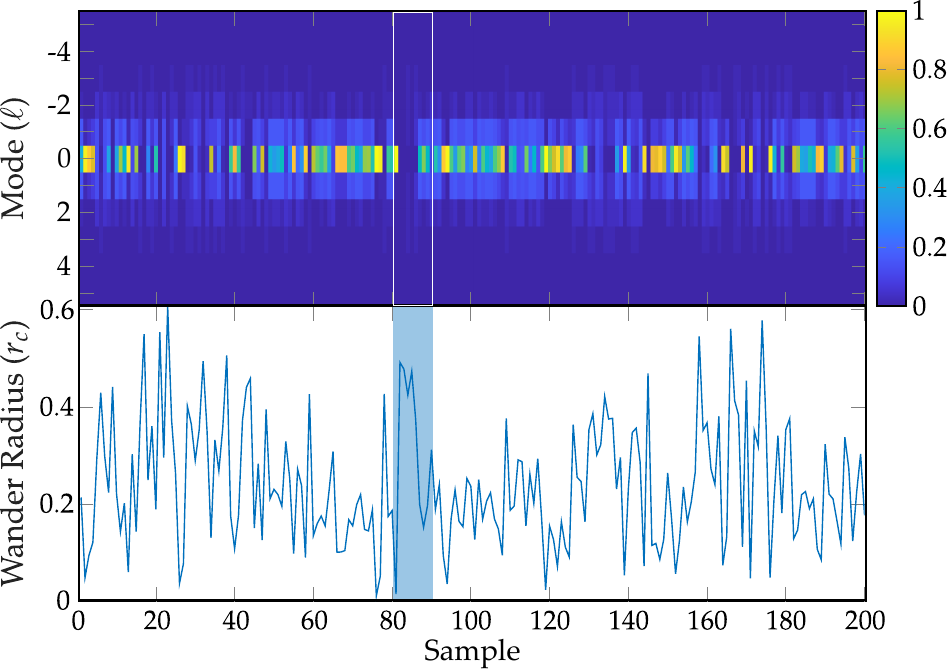}
	\caption{\label{fig:wanderCrosstalk} (Top) Mode crosstalk from the transmitted Gaussian mode ($\ell=0$) into OAM modes $\ell=-5$ to $5$ due to lateral displacement over time. (Bottom) The corresponding normalised beam wander radii, $r_c$. The highlighted section shows good examples of crosstalk with well aligned (minimal crosstalk) and misaligned (high crosstalk) cases. }
\end{figure}

The detected OAM spectrum due to a lateral displacement of $r_{c,t}$ at a time $t$ is given by \cite{Lin2010oamtilt}
\begin{equation}
\label{eq:modeCrosstalkWander}
    C_\ell = \exp\left(-\frac{r_{c,t}^2}{\omega_{\mathrm{ST}}^2} \right) I_{|\ell|}\left(\frac{r_{c,t}^2}{\omega_{\mathrm{ST}}^2} \right)
\end{equation}
where $C_\ell$ is the weight for the $\ell$th mode and $I_n(x)$ is the $n$th-order modified Bessel function of the first kind. Figure~\ref{fig:wanderCrosstalk} shows a sample of consecutive OAM crosstalk calculations from the simulated beam wander. Since we have shown a temporal correlation between beam wander positions, there will also be a similar temporal correlation between mode crosstalk spectrums. This may be harnessed in a predictive manner to implement more efficient signal processing strategies for MDM or modal diversity.

\section{Conclusion}
\label{sec:conclusion}

We have developed a new approach for modelling turbulence-induced beam wander and tested the approach on a model for a 150~m experimental link in weak-to-moderate turbulence. The resulting models are able to accurately reproduce the effects of memory in the atmospheric turbulence channel, where there is a temporal correlation in the movement of the beam. Existing, PDF-based models are not able to simultaneously model the fading probability distribution as well as the average time that the beam has wandered a certain radius from the receiver (one of the causes of deep fading) whereas this new memory model can. In addition to accurately modelling the received intensity distribution as well as the temporal evolution of the received intensity, we briefly show how the model may be used to simulate the evolution of crosstalk in MDM systems using OAM as an example. Our model will enable the development of optimised signal processing and error correction coding techniques for free-space optical communications which were not previously possible.



\ifCLASSOPTIONcaptionsoff
  \newpage
\fi



\bibliographystyle{IEEEtran}
\bibliography{bib.bib}
%


\end{document}